\documentclass[prd,letterpaper,tightenlines,nofootinbib,floatfix,showpacs]{revtex4}

\let\originalleft\left
\let\originalright\right

\renewcommand{\left}{\mathopen{}\mathclose\bgroup\originalleft}
\renewcommand{\right}{\aftergroup\egroup\originalright}

\usepackage{amssymb,graphicx}
\usepackage{amsmath}
\usepackage{comment}
\usepackage{enumitem}

\begin{document} 
\setlength{\belowdisplayskip}{8pt plus 2pt minus 5pt}
\setlength{\abovedisplayskip}{8pt plus 2pt minus 5pt}
\title{New, spherical solutions of non-relativistic, dissipative hydrodynamics}
\author{G. Kasza$^{1,2,3}$, L. P. Csernai$^{4,5}$, T. Csörgő$^{1,2}$}
\affiliation{$^1$Wigner RCP, H-1525 Budapest 114, POB 49, Hungary}
\affiliation{$^2$MATE Institute of Technology KRC, H-3200 Gy{\"o}ngy{\"o}s, M\'atrai \'ut 36, Hungary}
\affiliation{$^3$Eötvös Loránd University, H-1118 Budapest XI, P\'azm\'any P. 1/A,  Hungary}
\affiliation{$^4$University of Bergen, Allégt. 55, 5007 Bergen, Norway}
\affiliation{$^5$Frankfurt Institute for Advanced Studies, 60438 Frankfurt/Main, Germany}

\begin{abstract}
We present a new family of exact solutions of dissipative fireball hydrodynamics for arbitrary bulk and shear viscosities. The main property of these solutions is a spherically symmetric, Hubble flow field. The motivation of this paper is mostly academic: we apply non-relativistic kinematics for simplicity and clarity. In this limiting case, the theory is particularly clear: the non-relativistic Navier-Stokes equations describe the dissipation in a well-understood manner.  From the asymptotic analysis of our new exact solutions of dissipative fireball hydrodynamics, we could draw a surprising conclusion: this new class of exact solutions of non-relativistic dissipative hydrodynamics is asymptotically perfect.
\end{abstract}
\maketitle

\section{Introduction}
\label{sec:intro}
The nearly perfect fluid behaviour in high energy heavy ion collisions has been observed at Brookhaven National Laboratory’s Relativistic Heavy Ion Collider (RHIC) by the four RHIC collaborations, BRAHMS \cite{Arsene:2004fa}, PHENIX \cite{Adcox:2004mh}, PHOBOS \cite{Back:2004je}, and STAR \cite{Adams:2005dq}.
Currently, the main-stream of hydrodynamical modelling of high energy heavy ion reactions seems to  focus mostly on the numerical solutions of dissipative relativistic hydrodynamics. However, several analytic exact solutions are famous or very well known: the status of the field including applications of exact solutions of fireball hydrodynamics have been reviewed recently in ref.~\cite{deSouza:2015ena}, including a discussion of several of the open questions of this field.

On the topic of describing relativistic heavy ion collisions with exact solutions of hydrodynamics, the most well-known papers deal with perfect fluid hydrodynamics for a 1+1 dimensional, longitudinally expanding fireball, for example the boost-invariant Hwa-Bjorken solution~\cite{Hwa:1974gn,Bjorken:1982qr}, or an accelerating solution with finite rapidity density by Landau and Belenkij~\cite{Belenkij:1956cd}. One of the main focuses of seeking new exact solutions of relativistic hydrodynamics, is to generalize the Hwa-Bjorken solution to a 1+3 dimensional, accelerating flow  profile. 

One of the first famous 1+3 dimensional, spherically symmetric exact solutions of fireball hydrodynamics is
the Zim\'anyi-Bondorf-Garpman (ZBG) solution~\cite{Bondorf:1978kz}.
In the ZBG solution, a finite density and temperature profile is described, that both vanish after a time-dependent cutoff distance given by the scale parameter $R(t)$. The ZBG solution is accelerating and the flow-field corresponds to a spherically symmetric Hubble flow: $\mathbf{v}(\mathbf{r},t) = 
\frac{{\dot R(t)}}{R(t)} \mathbf{r} $. The ZBG solution has been generalized to a Gaussian density and a homogeneous temperature profile in ref.~\cite{Csizmadia:1998ef} and subsequently to an arbitrary temperature and a matching density profile in ref.~\cite{Csorgo:1998yk}, in both cases keeping the spherically symmetric Hubble flow field. The present manuscript generalizes these solutions for an arbitrary, temperature dependent speed of sound, using a thermodynamically consistent equation of state as proposed in ref.~\cite{Csorgo:2001xm}, and for a rather general, temperature dependent shear and bulk viscosities, using the non-relativistic Navier-Stokes equations. 
For the sake of simplicity and clarity, throughout the body of this work we assume the validity of a spherically symmetric Hubble flow.

In recent simulations of relativistic viscous hydrodynamics,   allowing the presence of small specific shear \cite{Romatschke:2007mq} and bulk \cite{Bozek:2017kxo} viscosities provides an acceptable description of the sQGP (strongly coupled quark-gluon plasma), so the importance of viscous corrections to the  Hwa-Bjorken solution and other exact solutions of relativistic perfect fluid hydrodynamics also emerges.

The effect of shear viscosity with longitudinal acceleration has been  investigated recently by a perturbative, relativistic solution of Navier-Stokes and Isreal-Stewart theory in ref.~\cite{Jiang:2020big}.
Dissipation in relativistic hydrodynamics have been addressed in an exact form as well by two recent manuscripts~\cite{Csanad:2019lcl, Csorgo:2020iug}. These works discuss analytic results on the effect of bulk viscosity on 1+3 dimensional exact solutions of the relativistic Navier-Stokes as well as the Israel-Stewart equations, generalizing the Hwa-Bjorken solution for non vanishing kinematic bulk viscosities in 1+3 dimensions \cite{Csanad:2019lcl, Csorgo:2020iug}. The velocity field of these solutions is also a Hubble-flow, which describes reasonably well the asymptotic velocity profile of  small and exploding fireballs, with possible applications even in high energy heavy ion collisions~\cite{Csorgo:2020iug}.

It seems to us that the new solutions of refs.~\cite{Csanad:2019lcl, Csorgo:2020iug} provide a great tool to investigate the late time effects of bulk viscosity in the hydrodynamic evolution. However, the fundamental equations of dissipative, relativistic hydrodynamics are debatable and so far several schemes have been proposed, but without a fundamental and generally accepted method, that are able to provide a stable and causal theory, and at the same time have also the correct non-relativistic limit: the non-relativistic Navier-Stokes equations. This motivated us to investigate the non-relativistic and spherically symmetric limit of the relativistic exact solutions described in refs. ~\cite{Csanad:2019lcl,Csorgo:2020iug}. It turned out that in this limit we obtained new, simple and exact solutions of the non-relativistic Navier-Stokes equations. This solution is described in this work. This solution  also allows for a straightforward generalization to a rotating and directional Hubble type flow-field. The discussion of these  non-spherical solutions is clearly beyond the scope of the present manuscript: we plan to present these solutions elsewhere.

In this paper, we present a new family of exact and analytic solutions of non-relativistic, dissipative hydrodynamics. Because of the non-relativistic regime, we do not seek for a direct application of these solutions to describe experimental data, although such applications are possible even in the perfect fluid limit, 
as demonstrated before in refs.~\cite{Bondorf:1978kz,Helgesson:1997zz}. In this work, we opted to remain in the entirely academic framework and do not compare our theoretical results to experimental data. Although we apply a spherically symmetric, 3 dimensional Hubble flow, the scale parameter $R = R(t)$ is a time dependent parameter, and its second derivative, hence the acceleration of the expansion is non-vanishing in some of the parametric solutions that we present. We also discuss coasting, accelerationless solutions for a homogenous pressure profile. 

The spherical symmetric Hubble profile excludes the possibility of the rotation of the fireball, and it also causes the cancellation  of the shear viscosity term from the dynamical equations of non-relativistic hydrodynamics. To simplify further the dissipative dynamics, we neglect possible heat conductivity effects. These simplifications allow us to focus on the main aim of this manuscript: to investigate the asymptotic effects of the bulk viscosity in the simplest possible case of fireball hydrodynamics.

\section{Navier-Stokes equations of non-relativistic, viscous hydrodynamics}
\label{sec:Navier-Stokes}
The dynamical equations of non-relativistic, viscous hydrodynamics
are discussed below.


The continuity equation of the conserved particle density $n$ reads as
\begin{equation}\label{eq:continuity}
\partial_t n + \nabla \left(n\mathbf{v}\right) = 0,
\end{equation}
where $n(\mathbf{r},t)\equiv n$ is a function of the spatial coordinates $\mathbf{r}=\left(r_x,r_y,r_z\right)$
and time $t$, while $\mathbf{v}(\mathbf{r},t)\equiv\mathbf{v}$  stands for the velocity field. We allow for a compressible expansion, $\nabla \mathbf{v} \neq 0$, keeping in mind that the fireball formed in the Little Bangs of heavy ion collisions corresponds to a quickly expanding fluid. 
We start from the Navier-Stokes equation for  the conservation of energy, which is expressed as
\begin{equation}\label{eq:energycons}
\begin{split}
    \partial_t \varepsilon + \nabla \left(\varepsilon \mathbf{v}\right) + p \nabla \mathbf{v}  = &\nabla \left(\lambda \nabla T \right) + \zeta \left(\nabla \mathbf{v}\right)^2 +
     2\eta \left[ \textnormal{Tr}\left(D^2\right)-\frac{1}{3}\left(\nabla\mathbf{v}\right)^2\right],
\end{split}
\end{equation}
where $\varepsilon(\mathbf{r},t)\equiv \varepsilon$ is the energy density, the pressure is denoted by $p(\mathbf{r},t)\equiv p$, $T(\mathbf{r},t)\equiv T$ is the temperature,  the heat conductivity coefficient is denoted by $\lambda$ while the shear and bulk viscosity coefficients are denoted by  $\eta$ and  $\zeta$,
respectively, and $D$ is Cauchy's strain tensor, which is defined as
\begin{equation}\label{eq:D-matrix}
D_{ik}=\frac{1}{2}\left(\frac{\partial v_i}{\partial r_k}+\frac{\partial v_k}{\partial r_i}\right).
\end{equation}
We consider the case of vanishing heat conduction, corresponding to the assumption of $\lambda \equiv 0$.
The Navier-Stokes form of the Euler equation stands for the momentum conservation:
\begin{equation}\label{eq:euler}
    mn\left(\partial_t+\mathbf{v}\nabla\right)\mathbf{v} +\nabla p = \nabla\left(\zeta \nabla\mathbf{v}\right) 
    + \eta \left[ \Delta \mathbf{v} + \frac{1}{3}\nabla\left(\nabla \mathbf{v}\right) \right],
\end{equation}
where the single particle mass is denoted by $m$. Here we assume that the bulk viscosity $\zeta$ may depend on position, so care must be taken when evaluating the first term on the right hand side of the above equation.

The microscopic properties of the flowing matter are its thermodynamical features. These are characterized by an equation of state, which closes the above set of partial differential equations of non-relativistic hydrodynamics. In this paper, we follow refs.~\cite{Csorgo:2001xm,Csanad:2012hr,Csorgo:2016ypf,Nagy:2016uiz,Csorgo:2018tsu} and assume that the equation of state can be written as
\begin{eqnarray}
    \varepsilon & = &  \kappa(T)p , \label{e:kappaT}\\
    p & = & n T. \label{e:pnT}
\end{eqnarray}
These equations of state are thermodynamically consistent~\cite{Csorgo:2001xm}. In the next section, we detail the temperature dependence of the speed of sound that follows from these equations of state. Although similar equations of state have been utilized in exact analytic solutions of both non-relativistic and relativistic hydrodynamics before, the corresponding temperature dependent speed of sound has not been described or detailed in earlier papers, as far as we know.

\section{Temperature dependence of the speed of sound}
\label{sec:cs(T)}

Now let us clarify the meaning of the $\kappa(T)$ function. The simplest case is if this function is a constant: if $\kappa(T)$ becomes independent of the temperature $T$, it is denoted as $\kappa$ and its relationship to the  adiabatic index $\gamma$ can be easily found as
\begin{equation}\label{eq:gamma-kappa}
    \gamma = \frac{C_p}{C_V} = 1+\frac{1}{\kappa}\:, 
\end{equation}
where $C_p$ is the heat capacity at constant pressure and $C_V$ stands for the heat capacity if the volume of the system is kept constant. This is also the reason why the adiabatic index $\gamma$
is also called the heat capacity ratio.

The well-known formula of the speed of sound reads as
\begin{equation}
    c_s^2(T,\mu)  = \left. \frac{\partial p}{\partial \epsilon}\right|_{\sigma/n},
\end{equation}
where $\sigma$ stands for the entropy density. In general, the speed of sound may thus be dependent on both the temperature and on the chemical potential of conserved charges.
In case of ideal gases as well as in case of our
selected equations of states, eqs.~(\ref{e:kappaT},\ref{e:pnT}),
the connection between $\kappa$ and the speed of sound can be obtained from eq.~\eqref{e:kappaT}, for a constant, temperature independent $\kappa$ as follows:
\begin{equation}\label{eq:cs-kappa}
    c_s^2(T)=\gamma \, \frac{T}{m}
    = \left(1+\frac{1}{\kappa}\right)\frac{T}{m}.
\end{equation}
In the case of a temperature dependent $\kappa(T)$ function, the formula for the temperature dependence of  speed of sound yields 
a more general expression:
\begin{equation}
    c_s^2(T) = \left[1+\left(\kappa(T) + T \frac{d\kappa(T)}{dT}\right)^{-1}\right]\frac{T}{m}. 
\end{equation}

Based on eqs.~\eqref{eq:gamma-kappa} and~\eqref{eq:cs-kappa} we can define a temperature dependent adiabatic index and we have found that $\gamma(T)$ still can be expressed as the ratio of generalized, temperature dependent heat capacities:
\begin{equation}\label{eq:gamma-kappa-T}
    \gamma(T)=\frac{C_p(T)}{C_V(T)} = 1+\left(\kappa(T) + T \frac{d\kappa(T)}{dT}\right)^{-1}.
\end{equation}
A further generalization is possible, for the case when  the particle mass depends on only the temperature. The derivation is similar and the  formulae remains nearly unchanged.
The speed of sound with a temperature dependent $\kappa(T)$ and particle mass $m(T)$ is obtained as:
\begin{equation} \label{eq:CsT-kappa-T}
    c_s^2(T) = \gamma(T) \frac{ T}{m(T)}. 
\end{equation}
We have also investigated the case of a multi-component matter which is composed of a mix of various hadrons, with  temperature dependent masses. In such a medium the speed of sound depends on the average mass of the hadrons instead of the single particle mass:
\begin{equation}\label{eq:cs_multi}
    c_s^2(T) = \gamma(T) \frac{T}{\langle m(T) \rangle}, 
\end{equation}
where
\begin{equation} 
    \langle m(T) \rangle = \frac{\sum\limits_i m_i(T) n_i(T)}{\sum\limits_i n_i(T)}.
\end{equation}
Eq.~\eqref{eq:cs_multi} is expressed in a general form for multi-component hadronic matter with temperature dependent $\kappa$ parameter and particle masses. It is clear that this form of  the speed of sound is independent of the temperature (in)dependence of both the adiabatic index (hence $\kappa$ as well) and the (average) particle mass: in the case of constant, temperature independent mass $m$ and $\kappa$ or adiabatic index $\gamma$, eq.~(\ref{eq:cs_multi})
reduces to the form for ideal gases, eq.~(\ref{eq:cs-kappa}).


\section{Scale and the continuity equations for spherically symmetric Hubble-flow}
\label{sec:spherical-Hubble}

We search for spherically symmetric, exact  solutions of the Navier-Stokes equations, where the velocity field is a Hubble flow with a time-dependent Hubble-parameter: 
\begin{equation}\label{eq:Hubble-flow}
    \mathbf{v} =
    \frac{\dot{R}}{R} \mathbf{r} \, = \, \frac{\dot{R}}{R}
    \left(r_x,r_y,r_z\right),
\end{equation}
where $R(t)\equiv R$ is the scale of the expanding fireball. In a Hubble flow, $R$ is a function of  the time, only. 
We seek self-similar solutions, thus we introduce a scaling variable  $s$ that satisfies the scale equation:
\begin{equation}
    \left(\partial_t + \mathbf{v}\nabla\right) s = 0.
\end{equation}
For the above defined spherically symmetric Hubble flow, a spherically symmetric solution of the scale equation is:
\begin{equation}
    s=\frac{r^2}{R^2},
\end{equation}
and the $\textbf{D}$ matrix, defined in eq.~\eqref{eq:D-matrix}, becomes  diagonal:
\begin{equation} \label{e:Hubble-flow-field}
    \textbf{D}=\frac{\dot{R}}{R}\:\textbf{I}.
\end{equation}
Using eq.~\eqref{eq:Hubble-flow} for the velocity field, the solution of the continuity equation is:
\begin{equation}\label{e:continuity-solution}
    n(\mathbf{r},t)=n_0\left(\frac{R_0}{R}\right)^d\mathcal{V}(s),
\end{equation}
where $\mathcal{V}$ is an arbitrary function of the $s$ scale variable, $n_0$ stands for $n(\mathbf{0},t_0)$, the initial value of the $R$ scale is $R_0\equiv R(t_0)$ and $d = 3$ is the number of spatial dimensions. 

The spherically symmetric Hubble flow-field given by eq.~(\ref{e:Hubble-flow-field}) and the spherically symmetric, generic solution for the density of the conserved charge, given by eq.~(\ref{e:continuity-solution}) will be common in various classes of exact solutions of the Navier-Stokes equations detalied in the subsequent parts of this manuscript.
Due to the spherical symmetry of  the selected velocity profile, the effects of shear viscosity cancel.

\section{New, exact solutions for a generic, temperature dependent speed of sound}\label{sec:solutions-with-T-dependent-cs}
In this section, we describe our general results, gradually introducing more and more simplifying assumptions. 
First of all, let us note that even in the $\kappa(T) = \kappa_0$
constant case, the considered equation of state leads to a temperature dependent speed of sound, as given in eq.~(\ref{eq:cs-kappa}). 
For a generic, temperature dependent pressure to energy density ratio $\kappa(T)$ the temperature dependence of the speed of sound may be even more complicated, see for example eq.~(\ref{eq:CsT-kappa-T}) and further  details in Section~\ref{sec:cs(T)}.
In this subsection, we collect results that are obtained for a generic,
temperature dependent pressure to energy density ratio $\kappa(T)$.

When searching for solutions with a temperature dependent $\kappa(T)$ function, we utilize earlier results of solving the equations of non-relativistic perfect fluid hydrodynamics with a temperature dependent $\kappa(T)$ and corresponding temperature dependent speed of sound, in particular the properties of the solutions in refs.~\cite{Csorgo:2001xm,Csorgo:2015scx,Nagy:2015bmt,Nagy:2016uiz,Csorgo:2016ypf,Csorgo:2018pxh}.
Based on these perfect fluid solutions, our ansatz for the self-similar temperature profile is the following,  factorized form:
\begin{equation}\label{eq:temp_ansatz}
    T(\mathbf{r},t)=T_0 f_T(t) \mathcal{T}(s),
\end{equation}
where $\mathcal{T}(s) \geq 0$ is an arbitrary non-negative function of $s$, normalized as $\mathcal{T}(0) = 1$. The non-trivial time dependence is described by the factor $f_T(t)$, and we denote the initial temperature at $s= 0$ at the initial time $t_0$ by $T_0$. For a non-zero conserved charge or bariochemical potential, the pressure is thus determined by the
equation of state, $p=n T$ as:
\begin{equation}
    p(\mathbf{r},t)=p_0 f_T(t) \left(\frac{R_0}{R}\right)^d\mathcal{V}(s)\mathcal{T}(s),
\end{equation}
where $p_0 = p(\mathbf{0},t_0) = n_0 \, T_0$. 
Using these considerations, the energy equation, eq.~(\ref{eq:energycons}) can be rewritten as follows:
\begin{equation}\label{eq:energycons_with_hubble-kappa-T}
    \frac{{\mathrm d}(T \kappa(T))}{{\mathrm d}T}\, \partial_t \ln f_T +d\frac{\dot{R}}{R}= d^2 \frac{\zeta }{p}\left(\frac{\dot{R}}{R}\right)^2.
\end{equation}
This equation can be further simplified and solved, if the temperature is spatially homogeneous, $\mathcal{T}(s) = 1$, keeping a generic $\kappa(T)$ function. For such a spatially homogeneous temperature profile,
the left hand side of the above equation depends on time, only.
For such a temperature dependent pressure to energy density ratio $\kappa(T)$, the temperature dependent adiabatic index $\gamma(T)$ is determined in Section~\ref{sec:cs(T)} as given by eq.~\eqref{eq:gamma-kappa-T} and the temperature dependent speed of sound by eq.~(\ref{eq:CsT-kappa-T}).

Eq.~\eqref{eq:energycons_with_hubble-kappa-T} is a non-linear but ordinary differential equation. Its left hand side depends on time, while its right hand side depends not only on time but also on the ratio of the bulk viscosity to pressure. Thus this equation can be solved if this ratio is assumed to be a constant:
\begin{equation}\label{e:zeta-per-p-const}
    \frac{\zeta}{p} = \frac{\zeta_0}{p_0} 
\end{equation}

The resulting equation, valid for spatially homogeneous temperature profiles, $\mathcal{T}(s) = 1$ reads as follows:
\begin{equation}\label{eq:energycons_with-kappa-T-zeta-p-const}
    \frac{{\mathrm d}(T \kappa(T))}{{\mathrm d}T}\, \partial_t \ln T +d\frac{\dot{R}}{R}= d^2 \frac{\zeta_0 }{p_0}\left(\frac{\dot{R}}{R}\right)^2.
\end{equation}

In the Euler equation the second derivatives cancel because of the special form of the velocity field, so we get back the same equation that corresponds to perfect fluids:
\begin{equation}\label{eq:euler-perfect}
    mn\left(\partial_t+\mathbf{v}\nabla\right)\mathbf{v} +\nabla p = \nabla\zeta\nabla\mathbf{v}.
\end{equation}
Using a spherical Hubble-profile and our ansatz for the temperature field, introduced by eq.~\eqref{eq:temp_ansatz}, the Euler equation is reduced to the following second order, ordinary differential equation: 
\begin{equation}\label{e:Euler-kappa-T}
    R\ddot{R} =  C_E\frac{T}{m} \left(1-3\frac{\zeta_0}{p_0}\frac{\dot{R}}{R}\right) = C_E\:f_T(t)\frac{T_0}{m} \left(1-3\frac{\zeta_0}{p_0}\frac{\dot{R}}{R}\right).
\end{equation}
Here a constant of integration is denoted by $C_E$, a constant parameter that also enters the following differential equation
for the scaling functions:
\begin{equation}\label{eq:diff_eq_of_scalefunc}
    -\frac{C_E}{2}=\mathcal{T}^\prime(s)+\frac{\mathcal{T}(s)}{\mathcal{V}(s)}\mathcal{V}^{\prime}(s),
\end{equation}
where the prime $\null^{\prime}$ in $\mathcal{V}^\prime(s)$ and $\mathcal{T}^\prime(s)$ denotes a derivation with respect to $s$. Eq.~\eqref{eq:diff_eq_of_scalefunc} indicates, that the $\mathcal{T}(s)$ and $\mathcal{V}(s)$ profile functions for the density and for the temperature  are not independent from each other: they must  be chosen corresponding to a matching initial condition.
The differential equation for the scaling functions can be integrated, and the scaling function for the density can be expressed with the help of the constant of integration and the scaling function of the temperature:
\begin{equation}\label{eq:scale_condition}
    \mathcal{V}(s)=\frac{1}{\mathcal{T}(s)}\exp\left(-\frac{C_E}{2}\int\limits_0^s \frac{du}{\mathcal{T}(u)}\right).
\end{equation}
It is thus clear that spatially homogeneous temperature profiles with
$\mathcal{T}(s) = 1$ correspond to Gaussian density profiles,
$\mathcal{V}(s) = \exp\left(-\frac{C_E}{2} s^2\right)$.

Thus, we have reduced the complex set of partial differential equations of non-relativistic, dissipative hydrodynamics with a temperature dependent pressure to energy ratio $\kappa(T)$ and the corresponding temperature dependent speed of sound given by 
eq.~(\ref{eq:CsT-kappa-T}), for spatially homogeneous temperature profiles,  to a family of coupled and non-linear, but ordinary differential equations,
eqs.~(\ref{eq:energycons_with_hubble-kappa-T},\ref{e:Euler-kappa-T}).
These equations can be easily solved with broadly accessible, numerical software packages like Matlab, Maple or Mathematica. Note that these solutions describe an accelerating fluid, where the second time derivative of the scale parameter, $\ddot R(t)$ is a non-vanishing function of time, as follows from eq.~(\ref{e:Euler-kappa-T}) for a non-vanishing constant of integration, $C_E \neq 0$. Accelerationless solutions are also obtained, according to eq.~(\ref{e:Euler-kappa-T}) they correspond to the $C_E = 0$ special case.

For the sake of completeness, let us also mention the equation for the $\sigma\equiv \sigma(\mathbf{r},t)$ entropy density, that describes the entropy production due to dissipation. This equation reads as:
\begin{equation}\label{eq:entropy_with_hubble}
    \partial_{t} \, \sigma + \frac{d}{t}\sigma = 
    \frac{\zeta}{T} \frac{d^2}{t^2} \ge 0.
\end{equation}

\section{Solutions for a temperature independent pressure to energy density ratio}
From now on, let us consider a temperature independent  energy density to pressure ratio $\kappa(T) = \kappa_0$. 
In this manuscript we are considering three dimensional, finite fireballs, however, such equations of states are frequently used, for example in the case of the conformal Gubser flows, that correspond to the $\kappa_0 = 3$ case as detailed in 
\cite{Gubser:2010ze, Gubser:2010ui}.
For such a temperature independent energy density to pressure ratio $\kappa_0$,  eq.~(\ref{eq:energycons_with_hubble-kappa-T})  simplifies as
\begin{equation}\label{eq:energycons_with_hubble}
    \kappa_0\, \partial_t \ln f_T +d\frac{\dot{R}}{R}= d^2 \frac{\zeta }{p}\left(\frac{\dot{R}}{R}\right)^2.
\end{equation}

\subsection{Analytic solutions for a spatially homogeneous pressure distribution}\label{subsec:solution_hompress}
We can provide exact analytic solutions of the above system of differential equations if the pressure distribution is spatially homogeneous. Although such a pressure distribution is academic, this solution is interesting as its certain properties may be inherited by solutions with spatially inhomogenous pressure profiles as well. 
Let us consider two important remarks. First, consider that in this case, when the pressure depends only on time, the coefficient of the bulk viscosity can be chosen to be any pressure dependent function because of the energy equation, eq.~\eqref{eq:energycons_with_hubble}:
\begin{equation}
    \zeta\equiv \zeta(p(t)).
\end{equation}
Second, the condition of homogeneity requires that $\mathcal{V}(s)\mathcal{T}(s)=1$. So according to eq.~\eqref{eq:scale_condition}, the constant of integration has to vanish, $C_E = 0$. Then the Euler equation is simplified to
\begin{equation}
    \ddot{R}=0.
\end{equation}
This equation reflects the fact that without a pressure gradient, the fireball expands with a constant rate, $\dot{R} = \dot{R}_0$.
Thus the time-dependent scale of the fireball is expressed by the function $R(t)=\dot{R} (t-t_0)$. Using this result, the energy equation is reduced to the following:
\begin{equation}\label{eq:energy_cons_homogen_pressure}
    \kappa_0\partial_t\textnormal{ln}\left(f_T\right)+\frac{d}{t}=\frac{\zeta(p(t))}{p(t)}\frac{d^2}{t^2}.
\end{equation}
If we assume, that the ratio of the bulk viscosity to pressure is a constant ($\zeta\propto p$), then we obtain an explicit solution for the $f_T(t)$ function:
\begin{equation}\label{eq:fT-hom}
    f_T(t)=\left(\frac{t_0}{t}\right)^{\frac{d}{\kappa_0}}\exp\left(\frac{d^2\zeta_0}{\kappa_0 p_0 t_0}\left[1-\frac{t_0}{t}\right]\right),
\end{equation}
where the ratio $\zeta_0/p_0$ is the proportionality factor between $\zeta$ and $p$. Utilizing this form, exact results can be obtained for the temperature, the conserved charge and the pressure:
\begin{align}  
      R(t) & = \dot{R}_0 t, \label{eq:Rt-hom-p} \\
    \mathbf{v} & = \frac{\dot R}{R} \mathbf{r} = \frac{\mathbf r}{t}, \label{eq:Hubble-without-acceleration}\\
    T(t,s) & = T_0 \left(\frac{t_0}{t}\right)^{\frac{d}{\kappa_0}}\exp\left(\frac{d^2\zeta_0}{\kappa_0 p_0 t_0}\left[1-\frac{t_0}{t}\right]\right)\mathcal{T}(s), \label{eq:temp_analytic_result} \\
    n(t,s) & = n_0 \left(\frac{t_0}{t}\right)^{d}
    \frac{1}{\mathcal{T}(s)}, \label{eq:n_analytic_result} \\
    p(t)&=p_0 \left(\frac{t_0}{t}\right)^{d\left(1+\frac{1}{\kappa_0}\right)}\exp\left(\frac{d^2\zeta_0}{\kappa_0 p_0 t_0}\left[1-\frac{t_0}{t}\right]\right). \label{eq:press_analytic_result}
\end{align}
This result is not only an analytic solution of the non-relativistic Navier-Stokes equations, but it turns out that this exact solution is the non-relativistic limit of a new, relativistic solution of dissipative hydrodynamics with accelerationless Hubble-profile, as detailed in ref.~\cite{Csorgo:2020iug}. Note that in eq.~\eqref{eq:Rt-hom-p}, an additional integration constant is allowed that controls the value of R(t) at $t_0$, but in this case we fixed this constant to 0.

\begin{figure}
    \centering
    \resizebox{0.45\textwidth}{!}{%
    \includegraphics{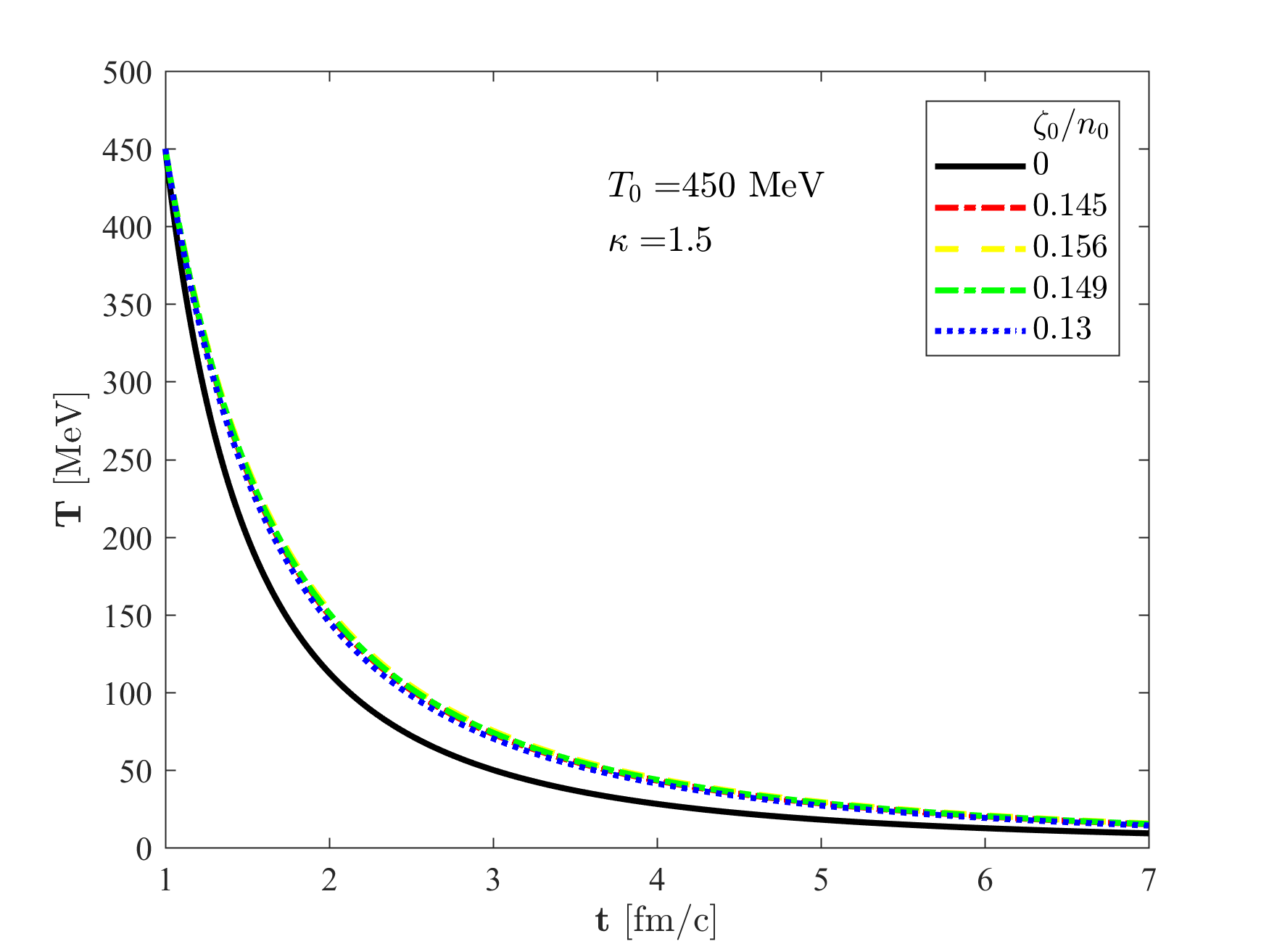}}
    \resizebox{0.45\textwidth}{!}{%
    \includegraphics{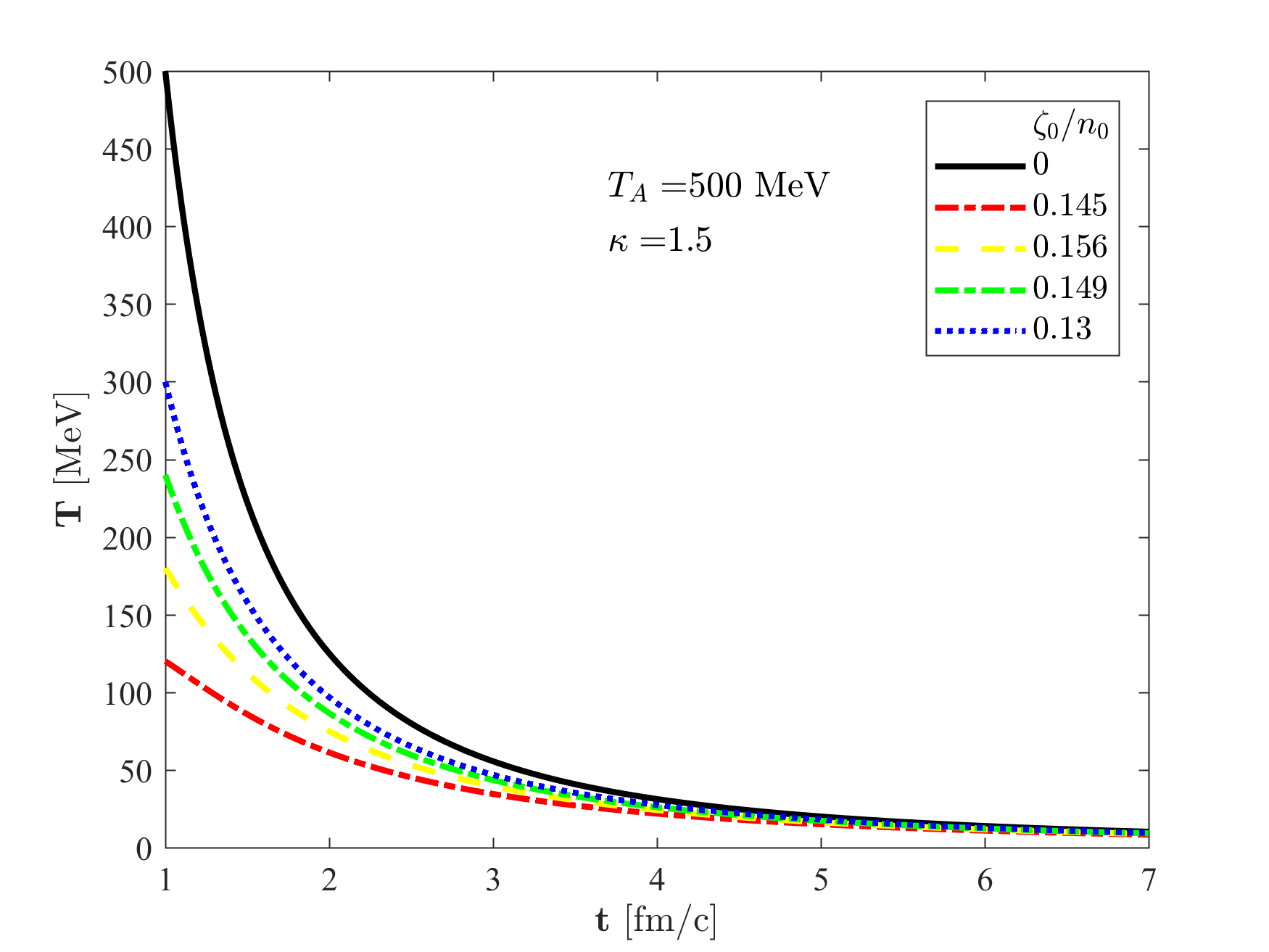}}
    \caption{The time evolution of the temperature in the center of the fireball ($s=0$). The solid black line corresponds to a perfect fluid solution, the colored lines correspond to our new, viscous solution of non-relativistic Navier-Stokes equation with homogeneous pressure field for different values of initial kinematic bulk viscosities. In the left panel we set the same initial temperatures, but in the right panel the curves start from different initial conditions and each of them approach the solid black line, the perfect fluid asymptote.}
    \label{fig:toymodel_temperature}
\end{figure}

This result allows for an asymptotic analysis, which is given in more details in subsection~\ref{subsec:attractor_hompress}. The key observation is coming from the analysis of the second factor of eq.~(\ref{eq:fT-hom}),
which tends to a constant multiplicative factor for late times, $t \gg t_0$. According to eq.~\eqref{eq:temp_ansatz} the time dependence of the temperature is governed by $f_T(t)$ which is illustrated in Fig.~\ref{fig:toymodel_temperature} for $s=0$. In the left panel we have shown five different curves with the same initial temperature and the initial kinematic bulk viscosity is varied. The effect of entropy production is clearly visible compared to the black curve which corresponds to perfect fluid. In the right panel we varied not only the initial value of kinematic bulk viscosity, but also the initial temperature. In this case each coloured curve tends to the same asymptote shown by a solid black line. Thus these curves approach the same final state. For a detailed explanation of this phenomenon see Section~\ref{asymptotically-perfect}.

\subsection{Analytic solutions for a spatially inhomogeneous pressure profile}\label{subsec:solution_inhompress}
Now let us discuss the case where the pressure also depends on the spatial coordinates through the scale variable $s$. In eq.~\eqref{eq:energycons_with_hubble} one can realize that the left side of the equation depends only on time, but the right side has coordinate dependence too, because both the pressure and the shear viscosity may depend on the scaling variable $s$. 
However, this  $s$ dependence of the right hand side is eliminated, 
if the bulk viscosity coefficient is proportional to the pressure.
Let us assume, that the bulk viscosity has a special form:
\begin{equation}
    \zeta(t,s)=\zeta_0 \frac{p(t,s)}{p_0}.
\end{equation}
In this case, the energy equation, eq.~\eqref{eq:energycons_with_hubble} becomes an ordinary differential equation that depends on time only. 
In addition, let us factor out a trivial time dependence that is due to the decrease of the temperature due to the radial expansion.  Let us rewrite  the time dependent factor $f_T(t)$ in the following form:
\begin{equation}\label{e:gT}
    f_T(t)=g_T(t)\left(\frac{R_0}{R}\right)^\frac{d}{\kappa_0},
\end{equation}
where $g_T(t)$ stands for a dissipative correction in the time dependence of the temperature field. For a vanishing bulk viscosity coefficient, $g_T(t) \equiv 1$. This ansatz is based on known non-relativistic, perfect fluid solutions with Hubble-flow, as detailed in refs.~\cite{Csizmadia:1998ef, Akkelin:2000ex, Csorgo:2001xm, Csorgo:2002kt, Csorgo:2013ksa, Nagy:2016uiz}. Using this ansatz, 
the energy equation of eq.~(\ref{eq:energycons_with_hubble})  can be rewritten in terms of $g_T(t)$ and $R(t)$ as follows:
\begin{equation}\label{eq:energy_with_gT}
    \frac{\dot{g}_T}{g_T} = 
    \frac{\zeta_0 d^2} { p_0 \kappa_0}\left(\frac{\dot{R}}{R}\right)^2.
\end{equation}
The Euler equation can be also rewritten in terms of the  dissipative correction $g_T$ as follows:
\begin{equation}\label{eq:euler_with_gT}
    R\ddot{R}=C_E\:g_T(t) \frac{T_0}{m}\left(\frac{R_0}{R}\right)^\frac{d}{\kappa_0}\left(1-3\frac{\zeta_0}{p_0}\frac{\dot{R}}{R}\right).
\end{equation}
Equations (\ref{eq:energy_with_gT},\ref{eq:euler_with_gT}) are time-dependent, coupled and non-linear, but ordinary differential equations for $R(t)$ and $g_T(t)$. These equations can be readily solved with the help of generally available mathematical packages like Maple, Matlab or Mathematica. Furthermore, their structure is particularly clear and allows for an asymptotic analysis, that corresponts to their late time, $t \gg t_0$ behaviour. This is the subject of the analysis of the next section.

\subsection{Discussion: attractor behaviour in other hydrodynamical solutions}
\label{attractor-perfect}
Before we discuss the asymptotic properties of our newly described exact viscous solutions, let us mention that these asymptotic properties have a kind of attractor behaviour. Finding attractors in hydrodynamic systems as well as finding hydrodynamical attractors in non-equilibrium systems is a broad topic of great current research interest, that is impossible to fully review in the current, rather academic manuscript with more limited scope.

In ref.~\cite{Dash:2020zqx}, the Boltzmann equation was solved in a relaxation time approximation, for an azimuthally symmetric radially expanding boost-invariant conformal system that is undergoing a longitudinally boost-invariant, radial Gubser flow~\cite{Gubser:2010ui}, and hydrodynamical attractor solution was found in  various approximations of relativistic viscous hydrodynamics.
Analytic solutions of dissipative relativistic spin hydrodynamics were studied recently based on a  Gubser expansion. 
The evolution of spin potential turned out to be important in the future studies of spin polarization in these solutions~\cite{Wang:2021wqq}.
There were several other important studies of dissipative relativistic hydrodynamics based on the perturbations or other modifications of the 
relativistic and boost-invariant, hence infinite Gubser flow. 
Instead of detailing these results, due to the limited focus of our work, let us quote some recent review papers on these topics.
For a very recent and rather pedagogical introduction to the effective descriptions relevant for attractors, in particular hydrodynamical attractors in high energy heavy ion physics, holography and kinetic theory, together by highlights of some recent advances in dissipative hydrodynamics we recommend ref.~\cite{Soloviev:2021lhs}. The status of dissipative relativistic hydrodynamics has been reviewed less recently, but more extensively in refs.~\cite{Romatschke:2017ejr} and~\cite{DerradideSouza:2015kpt}. This latter review~\cite{DerradideSouza:2015kpt}, discussed and presented
also the topics of exact solutions of perfect fluid hydrodynamics, so relevant background for our current manuscript.

Indeed, for the context of our exact dissipative solutions and their relation to perfect fluid solutions, 
it is useful to mention, that some exact non-relativistic as well as relativistic perfect fluid solutions seem to form a basis or a background solution, that are recovered by our results in the limit of vanishing dissipation.
For example, our results for the special case of $\kappa = \kappa_0$ generalize the
Zim\'anyi-Bondorf-Garpman~\cite{Bondorf:1978kz} spherically symmetric perfect fluid solutions for non-vanishing bulk viscosity and for a cancelling shear viscosity coefficient, as evident from the scale equation: substituting $\zeta_0 = 0$ to eqs.~(\ref{eq:euler_with_gT},\ref{eq:energy_with_gT}) one recoveres the scale equation for $R(t)$ obtained already in ref.~\cite{Bondorf:1978kz}.

The scaling function of the density as well as for the temperature profile are related with a matching initial condition, eq.~(\ref{eq:scale_condition}). In ref.~\cite{Csorgo:1998yk}, this equation was also obtained before but for a perfect fluid and a corresponding perfect fluid interpretation of the constant of integration (denoted by $C_E$ in the present manuscript and by $C_{\phi}$ ref.~\cite{Csorgo:1998yk}).
It is also inspiring to note that if this constant of integration is vanishing, $C_E = 0$, we obtain an inverse relationship between the scaling functions of the density and the temperature profiles, $\mathcal{V}(s)\mathcal{T}(s)=1$. This matching condition is not limited to the non-relativistic kinematic domain, as it is found to be valid also for relativistic perfect fluid solutions as detailed in ref.~\cite{Csorgo:2003ry}.

It is also inspiring to note that the above mentioned class of perfect fluid solutions has a straightforward generalization to spheroidal and ellipsoidal flows. The first triaxial, finite fireball solution, as far as we know, has been found by De, Garpman, Sperber, Bondorf and Zim\'anyi~\cite{De:1978iat}. This oblate, ellipsoidal perfect fluid solution has been generalized to an arbitraty positive definite temperature profiles in refs.~\cite{Csorgo:2001ru,Csorgo:2001xm}. This suggests that a generalization of the spherically symmetric, viscous solutions to oblate shaped, ellipsoidal fireballs may become feasible. 

Furthermore, these expanding  spheroidal, or triaxial ellipsoids may not only expand and cool, but rotate too. Such kind of exact and rotating fireball solutions have been obtained for non-relativistic perfect fluids in refs.~\cite{Csorgo:2013ksa,Csernai:2014hva,Csernai:2015jsa,Kasza:2018qah}. Due to the limited scope of this manuscript we do not provide a complete list of references on exact rotating solutions of fireball hydrodynamics, as the already quoted papers, in particular ref.~\cite{Kasza:2018qah} seem to be sufficient to conjecture, that viscous corrections of the non-relativistic Navier-Stokes equations can be obtained even for rotating, triaxially expanding ellipsoidal fireballs too.

Thus there seems to be a very interesting and deep connection between perfect fluid solutions and dissipative solutions of the Navier-Stokes equations. In the next section, we detail their relationship for the spherically symmetric, non-relativistic flows, the main topic of the current manuscript.

\section{Asymptotically perfect fluid behaviour}
\label{asymptotically-perfect}

During the understanding of these new solutions we have found that at late times, an expanding fireball with Hubble-type velocity profile is proceeding towards ``perfection", or in other words these new exact solutions of dissipative hydrodynamics are asymptotically equal to an exact solution of a perfect fluid hydrodynamics in the following well-defined sense: two functions of time $t$, called $x(t)$ and $y(t)$ are asymptotically equal, if 
\begin{equation}
    \lim_{t\rightarrow \infty} \frac{x(t)}{y(t)} = 1.
\end{equation}
Such an asymptotic equality is denoted by $x \sim y$ throughout this manuscript.

The asymptotic equality of a dissipative solution of hydrodynamics with a perfect fluid solution implies, that the effect of bulk viscosity can be scaled out at late times, corresponding  to low temperatures. 
This result supports the conclusion of \cite{Csorgo:2020iug}, that was obtained for relativistic kinematics, for a similar set of equations of state. In other words, this result implies, that we cannot decide from final state measurements that the medium evolved as a perfect fluid with higher initial temperature and entropy content, or as a viscous fluid with lower initial temperature. This behaviour is also perceived in ref.~\cite{Csanad:2019lcl}, that as far as we know presents the first exact solutions of the relativistic Navier-Stokes equations for non-zero bulk viscosities.

A similar, but different phenomenon is reported in ref.~\cite{Blaizot:2020gql}, which presents an analytic  solution of a set of differential equations that describe the transition from kinetic theory to hydrodynamics in the Bjorken expansion. Another interesting feature is provided by ref.~\cite{Janik:2005zt}: for large times, a unique, non singular, asymptotic solution of the non-linear Einstein equations in the bulk is found to be a perfect fluid solution of relativistic hydrodynamics.
The conclusion of refs.~\cite{Blaizot:2020gql,Janik:2005zt} also highlight the importance of the identification of attractor solutions in dynamical equations and the theory of asymptotic series.

In the following subsection we show how the approach to ``perfection" can be described with analytic tools in the simplest case. Subsequently,  we discuss this behaviour in more general cases and illustrate such a behaviour in two figures as well.

\subsection{Asymptotic analysis for the case of a homogeneous pressure and linear bulk viscosity}\label{subsec:attractor_hompress}
We have presented a new family of solutions with homogeneous pressure and arbitrary, pressure dependent bulk viscosity in subsection~\ref{subsec:solution_hompress}. At the end of that section, we found an analytic form for the time-dependent part of the temperature profile, if we  considered the case of $\zeta \propto p$,
namely a bulk viscosity coefficient that is a linear function of the pressure:
\begin{equation}
    \zeta(p(t))=\frac{\zeta_0}{p_0} p(t),
\end{equation}
where $\zeta_0=\zeta(p_0)$. In this case, the analytic form of the temperature, the conserved charge density, and the pressure has been obtained in eqs.~(\ref{eq:temp_analytic_result}-\ref{eq:press_analytic_result}). 
For late times, $t \gg t_0$, the exponential terms in the $T(t,s)$ temperature and the $p(t)$ pressure are approaching a constant value. Thus these fields are asymptotically are given as follows:
\begin{align}
    T(t) &\sim T^A_0 \left(\frac{t_0}{t}\right)^{\frac{d}{\kappa_0}}\mathcal{T}(s),\\
    p(t) &\sim p^A_0 \left(\frac{t_0}{t}\right)^{d\left(1+\frac{1}{\kappa_0}\right)}.
\end{align}
where we stress that the $\sim$ sign stands for the asymptotic equality. The initial temperature and initial pressure are rescaled by the asymptotic dissipative correction and absorbed by the asymptotic perfect fluid initial temperature ($T^A_0$) and pressure ($p^A_0$):
\begin{align}
    T^A_0 &= T_0\exp\left(\frac{d^2\zeta_0}{\kappa_0 p_0 t_0}\right), \label{eq:TA} \\
    p^A_0 &= p_0\exp\left(\frac{d^2\zeta_0}{\kappa_0 p_0 t_0}\right).
\end{align}
The effect of bulk viscosity is also absorbed into these parameters, so the bulk viscosity effect can be absorbed into the normalization of the initial pressure and temperature, and  has no other asymptotic influence on the time-evolution of the fireball. Accordingly, this asymptotic limit corresponds to a perfect fluid fireball hydrodynamics with $T(t_0)=T^A_0$ and $p(t_0)=p^A_0$ initial conditions. This family of perfect fluid solutions is already known, and published in ref.~\cite{Csorgo:2013ksa}. However, that reference discusses a wider family of perfect fluid solutions: the spherically symmetric fireballs are generalized to spheroidally symmetric, rotating fireballs.

\begin{figure}
    \centering
    \resizebox{0.7\textwidth}{!}{%
    \includegraphics{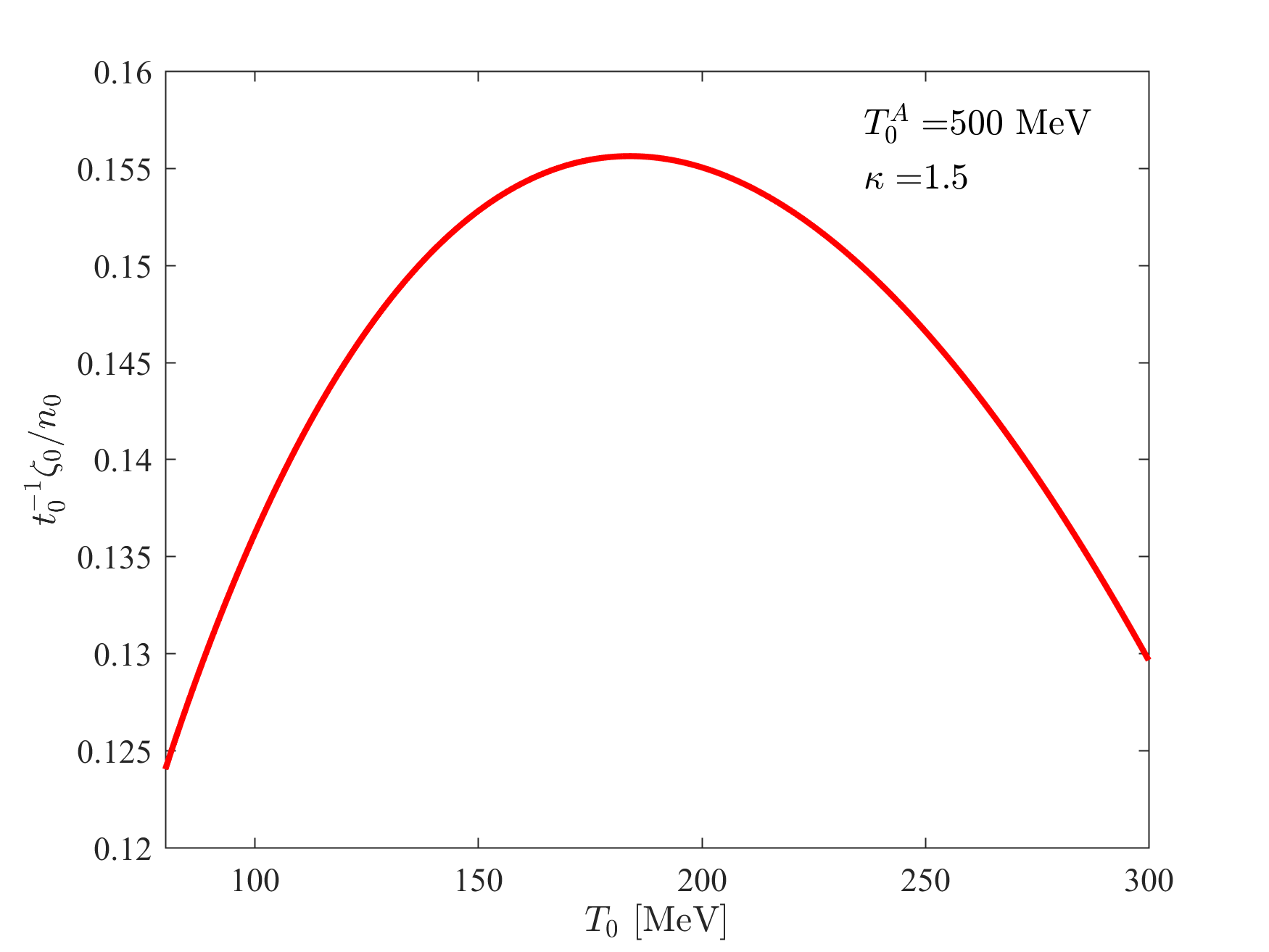}}    
    \caption{For a fixed asymptotic solution with fixed $T^A_0$, the initial value of the kinematic bulk viscosity is a function of $T_0$ and in this figure the initial time is scaled out. This non-monotonic behaviour is described by eq.~\eqref{eq:TA} and the maximum of the curve is given by eq.~\eqref{eq:max_of_zeta0ton0}.}
    \label{fig:zeta0ton0_and_TAtot0}
\end{figure}

An interesting property of the kinematic bulk viscosity ($\zeta_0/n_0$) emerges from eq.~\eqref{eq:TA}. Using the $p_0=n_0T_0$ ideal gas approach one can realize that the initial value of the kinematic bulk viscosity is a non-monotonic function of $T_0$:
\begin{equation}
    \frac{\zeta_0}{n_0}=\frac{\kappa_0 T_0 t_0}{d^2}\ln\left(\frac{T^A_0}{T_0}\right),
\end{equation}
and this function is illustrated in Fig.~\ref{fig:zeta0ton0_and_TAtot0}, where $\zeta_0/n_0$ is divided by the unobservable initial time $t_0$. Eq.~\eqref{eq:TA} sets an upper limit on $\zeta_0/n_0$ and it is written as:
\begin{equation}
    \frac{\zeta_0}{n_0} \le \frac{\kappa_0 t_0}{d^2} \frac{T^A_0}{e}, \label{eq:max_of_zeta0ton0}
\end{equation}
where $e$ denotes the Euler number. $T_A$ is an observable quantity by describing the experimental data with the perfect fluid solution of ref.~\cite{Csorgo:2013ksa}.

\subsection{Asymptotic analysis for an inhomogeneous pressure and linear bulk viscosity}
\label{subsec:asymptotic}
\begin{figure}[ht]
    \centering
    \resizebox{0.49\textwidth}{!}{%
    \includegraphics{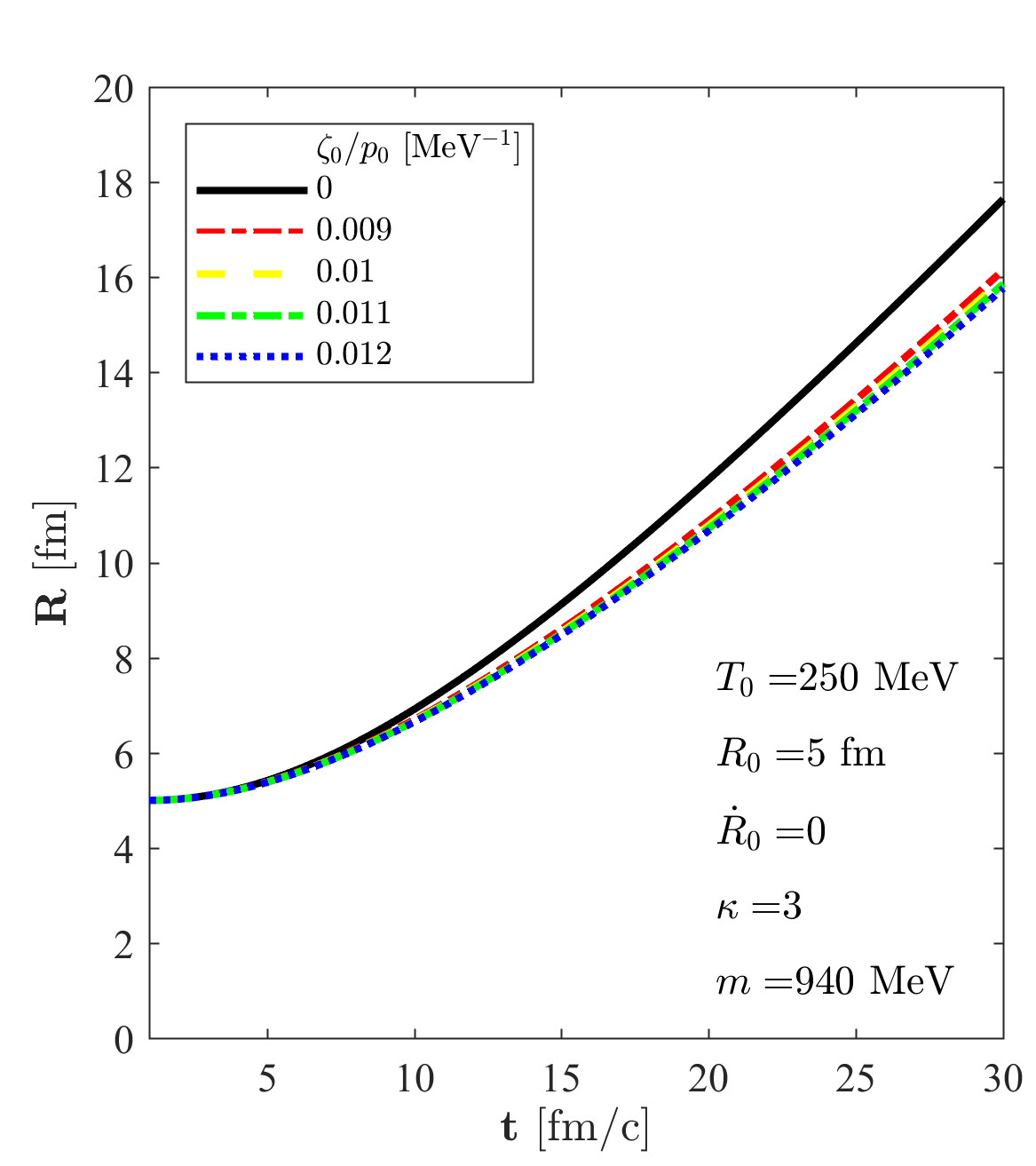}}
    \resizebox{0.49\textwidth}{!}{%
    \includegraphics{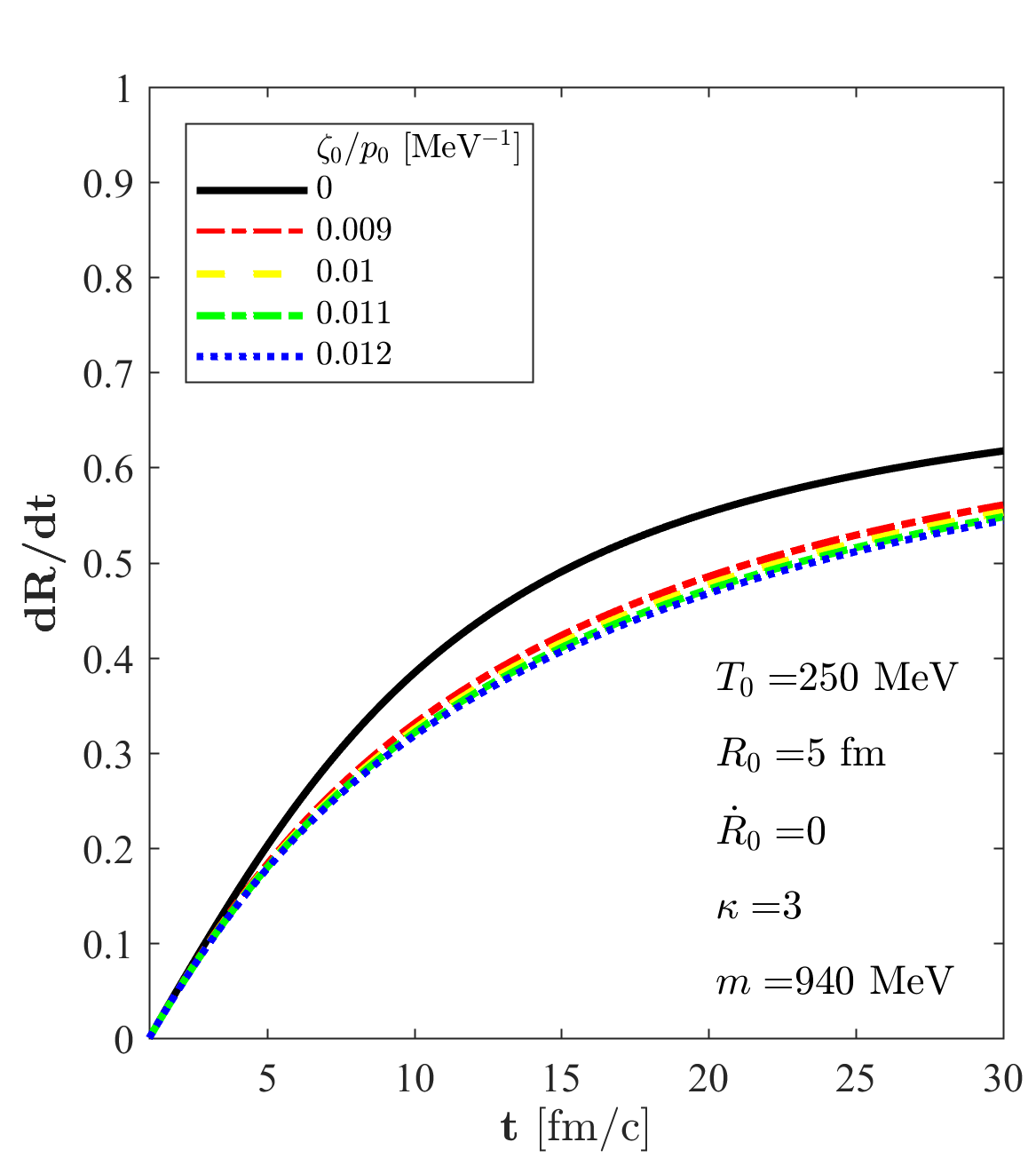}}
    \resizebox{0.49\textwidth}{!}{%
    \includegraphics{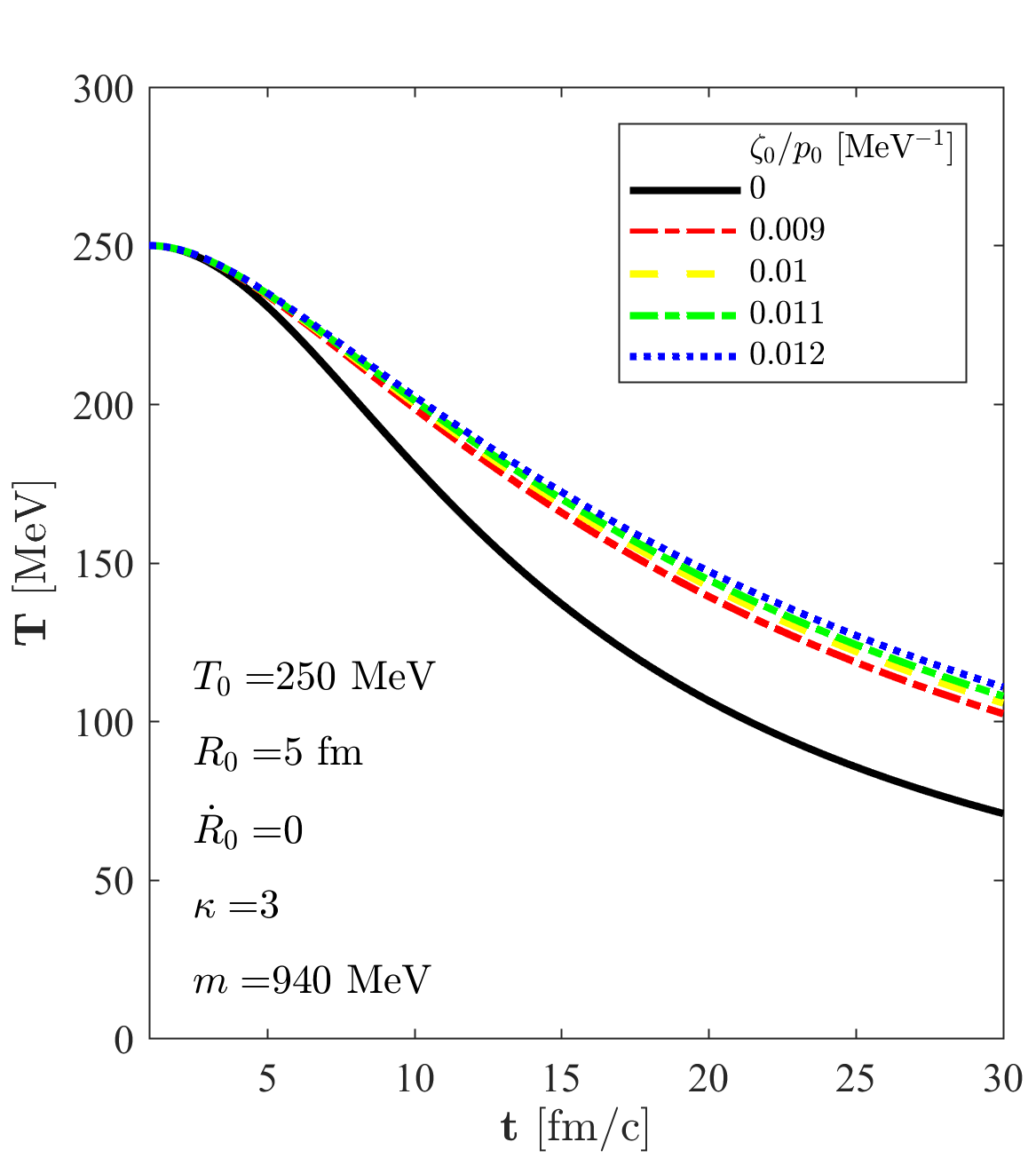}}    
    \caption{The evolution of the $R(t)$ scale of the fireball (upper left), and its time derivative $\dot{R}(t)$ (upper right), and the temperature (bottom) as a function of time for an exact solution of the non-relativistic Navier-Stokes equations for fixed $T_0=250$ MeV, $R_0=5$ fm and $\dot{R}_0=0$ initial parameters. We assume a nuclear fluid here with $m=940$ MeV particle mass and a constant, temperature independent $\kappa$ parameter: $\kappa = 3$. The solid black line stands for a perfect fluid solution, the dashed blue, the dotted-dashed green, the dashed yellow and the dotted-dashed red lines correspond to our new viscous solution of non-relativistic Navier-Stokes equations for different values of $\zeta_0/p_0$, but for the same initial conditions.}
    \label{fig:fixed_initial_conditions}
\end{figure}

\begin{figure}[ht]
    \centering
    \resizebox{0.49\textwidth}{!}{%
    \includegraphics{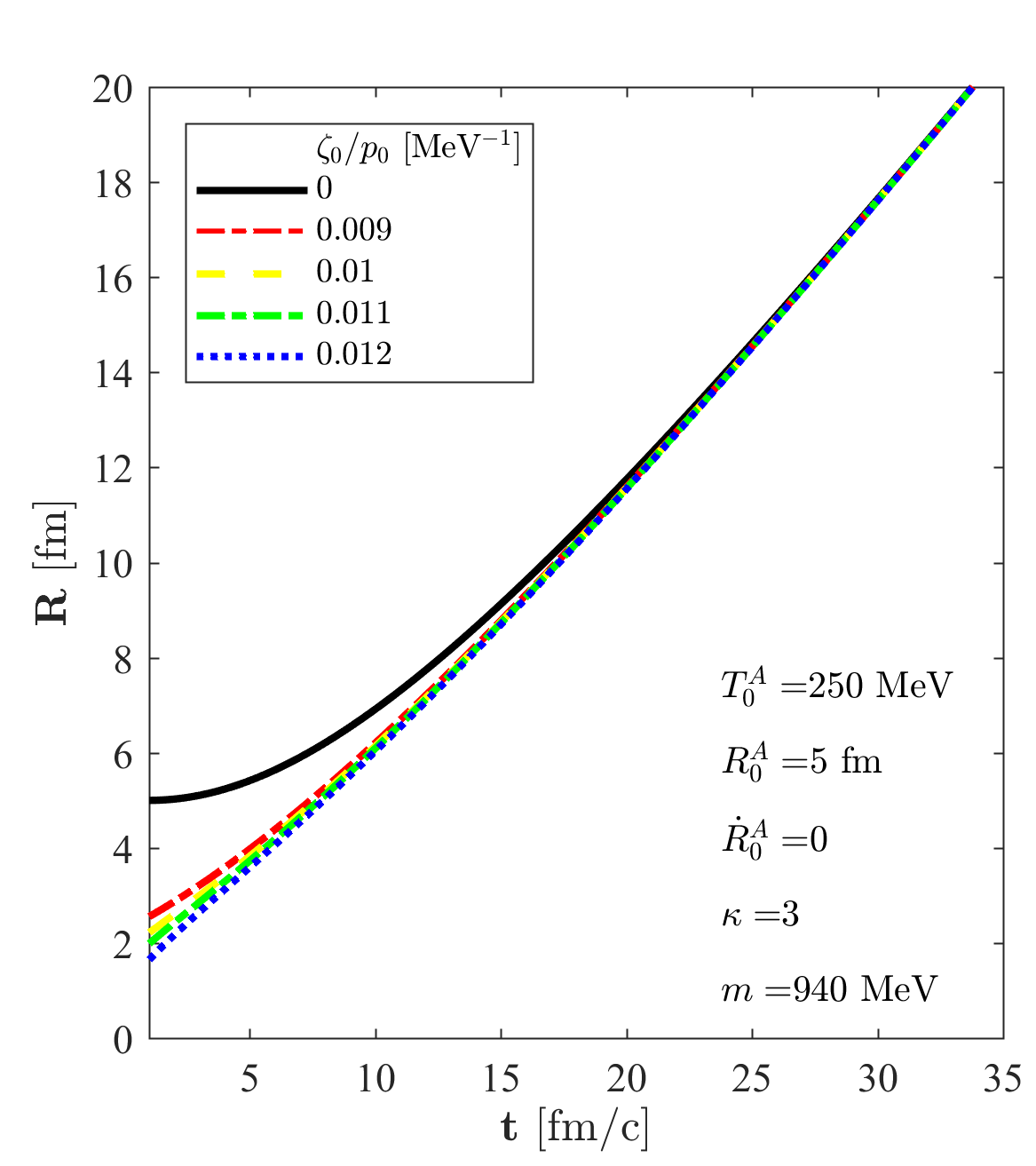}}
    \resizebox{0.49\textwidth}{!}{%
    \includegraphics{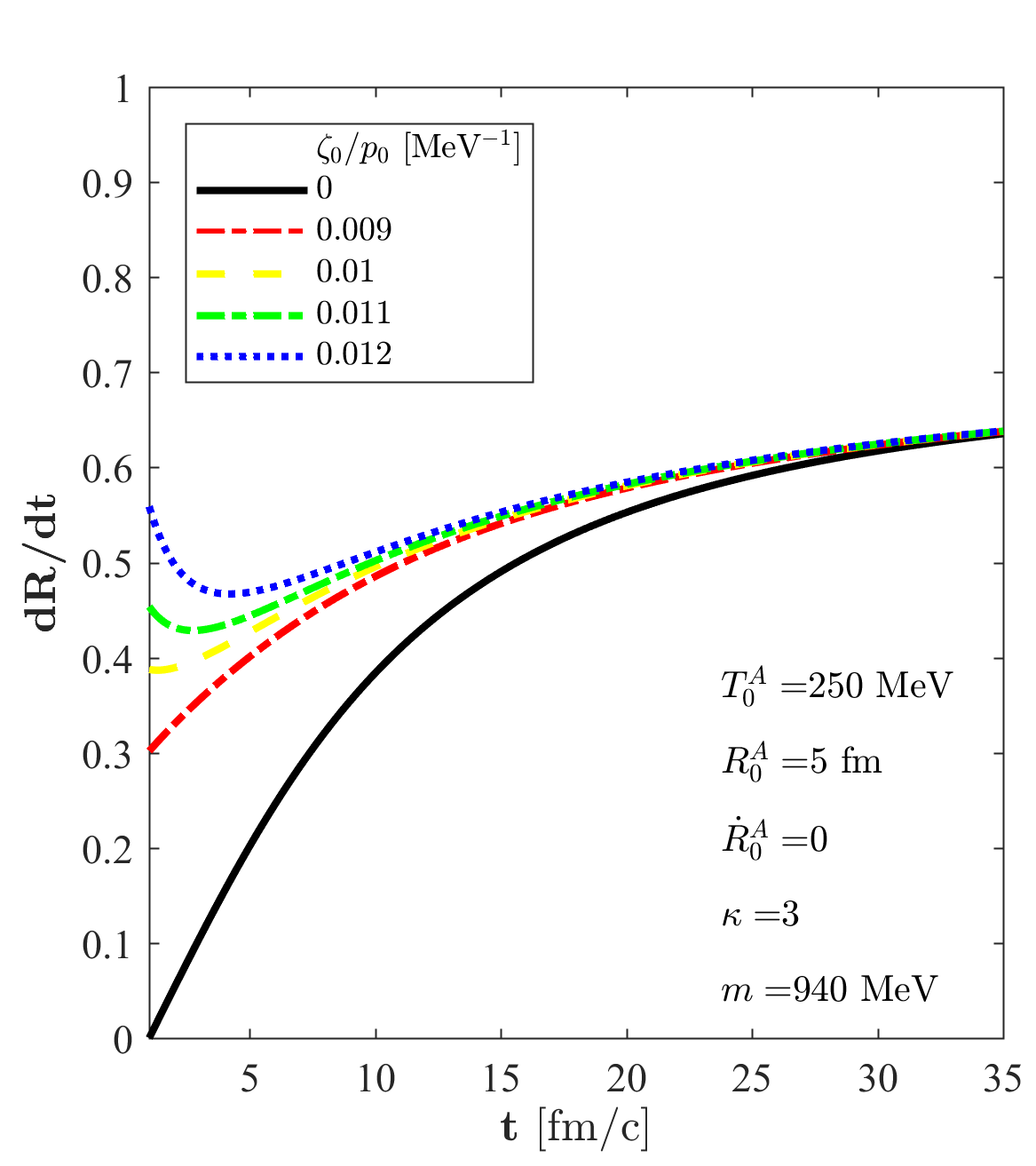}}
    \resizebox{0.49\textwidth}{!}{%
    \includegraphics{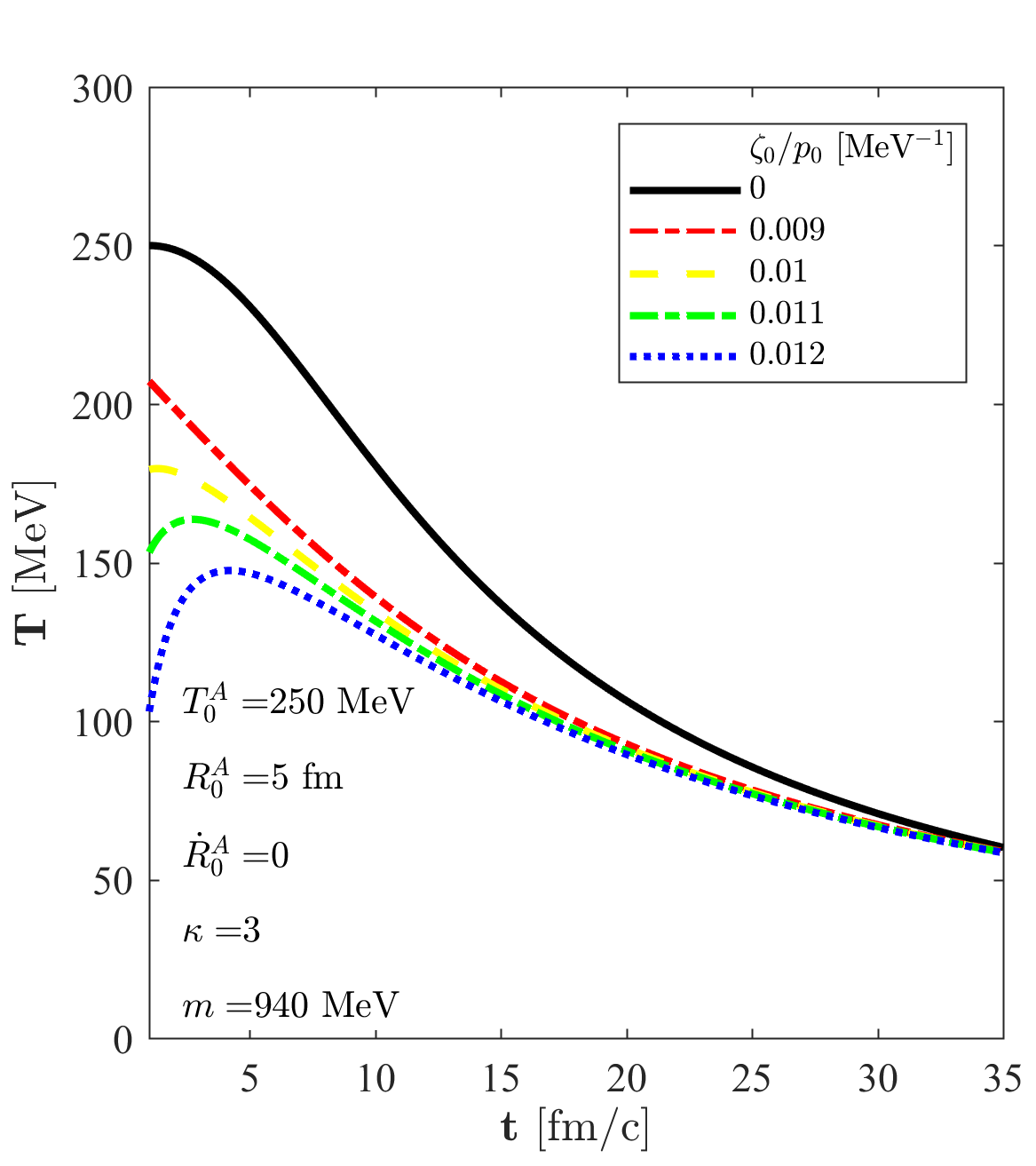}} 
    \caption{The evolution of the $R(t)$ scale of the fireball (upper left), the $\dot{R}(t)$ scale velocity (upper right), and the temperature (bottom) as a function of time for the solution of the non-relativistic Navier-Stokes equations for $T_0^A=250$ MeV, $R_0^A=5$ fm and $\dot{R}_0^A=0$ initial parameters, utilizing an $m=940$ MeV for the particle mass and a constant, temperature independent $\kappa = 3$. The solid black line stands for a perfect fluid solution, and this perfect fluid curve labelled by zero bulk viscosity is  approached by each of the shown exact viscous solutions asymptotically,  $T(t) \sim T_A(t)$. The dashed blue, the dotted-dashed green, the dashed yellow and the dotted-dashed red lines correspond to our new viscous solution of non-relativistic Navier-Stokes equations for different values of $\zeta_0/p_0$, but for the same asymptotic solutions.}
    \label{fig:fixed_finalstate_conditions}
\end{figure}

Now, let us examine the asymptotic behaviour of the more general solution presented in subsection~\ref{subsec:solution_inhompress}. It is a parametric solution, described by two coupled differential equations, namely eqs.~\eqref{eq:energy_with_gT} and~\eqref{eq:euler_with_gT}. We have numerically solved this set of differential equations for a perfect fluid and also for viscous fluids with different values of $\zeta_0/p_0$. We obtained the $R(t)$ scale of the fireball and the $g_T(t)$ function of the dissipative correction of temperature. These calculations are performed twice by two different approaches. The first approach is the conventional one, when we start the hydro evolution from the same initial conditions, and vary the amount of bulk viscosity. The results of these calculations  are shown in Fig.~\ref{fig:fixed_initial_conditions}, where one can see that the heating effect of the entropy production and that the same initial conditions lead to different asymptotic final states. This is a conventional feature of our solution. The second approach is when the initial conditions are varied, but the asymptotic equality of the solutions with different bulk viscosity coefficients are required. This is a non-conventional approach. We illustrated these cases in Fig.~\ref{fig:fixed_finalstate_conditions}. Looking at this figure one can realize that the black curves, related to perfect fluid solutions, behave as asymptotic attractors. Thus this figure provides the same conclusion that we have shown in the previous subsection for the case of homogeneous pressure with analytical tools: the exact solutions of dissipative fireball hydrodynamics are asymptotically equal to a perfect fluid solution with increased initial temperature and entropy content. If the asymptotic perfect fluid solution is fixed, then varying the bulk viscosity coefficients can be  co-varied with the initial conditions, so that asymptotically the same perfect fluid solution governs or attracts the late time behaviour of these dissipative solutions. In Fig.~\ref{fig:fixed_finalstate_conditions}, we characterized the curves with the particle mass $m$, the energy density to pressure ratio $\kappa$ that characterizes the equations of state, and the initial properties of the asymptotic perfect fluid attractor. We find that only the following physical quantities are important in determining the physical characteristics of the perfect fluid asymptotic solution: the  initial temperature of the asymptotic perfect fluid solution $T_0^A$,  the initial value of the scale $R_0^A$ and the initial velocity of the scale $\dot{R}_0^A$. These quantities, together with the initial density $n_0$ and the particle mass $m$ determine a perfect fluid solution that becomes asymptotically equal with the dissipative exact solution of the non-relativistic Navier-Stokes equations.


\section{Summary}
\label{sec:summary}
We have found a new family of analytic and parametric, exact solutions of non-relativistic Navier-Stokes hydrodynamics in 1+3 dimensions. In these solutions the velocity field is spherically symmetric Hubble-flow, so the effect of shear viscosity cancels, and we assumed that the heat conduction is negligible. With these features we provide exact and analytic solutions for the cases of homogeneous and inhomogeneous pressure as  well. 
Let us comment that even ignoring heat flow and assuming that the temperature and pressure are homogeneous is still appropriate for analysis of non relativistic, dissipative hydrodynamics, and even for the analysis of non relativistic heavy ion collisions as shown for example in refs.~\cite{Bondorf:1978kz,Helgesson:1997zz}. 
The solutions for homogeneous pressures correspond to the non-relativistic limit of a recently found exact solution of the relativistic Navier-Stokes equation \cite{Csorgo:2020iug}. This result strengthens the conclusion of refs.~\cite{Csorgo:2020iug,Csanad:2019lcl}. In this class of exact solutions of dissipative hydrodynamics,  for asymptotically late times, the effects of bulk and shear viscosity coefficients can be scaled out
and asymptotically the same hadronic final state or attractor solution can be reached using different bulk and shear viscosity coefficients, with suitably co-varied initial conditions. These solutions share a common perfect fluid asymptotic solution.

The non relativistic kinematic domain may make the significance of this paper rather academic. On one hand,  the assumption of spherical symmetry, the absence of heat flow and assuming homogeneous pressure and temperature field may obliterate the physical properties of dissipative fluids. On the other hand, we have found the effect of approach to ``perfection" first in the  relativistic kinematic domain~\cite{Csanad:2019lcl,Csorgo:2020iug}, before writing it up in the current, clear-cut non-relativistic form.
Although the domain of applicability is limited to spherical symmetry in the present manuscript, we conjecture that similar process of asymptotic  ``perfection" of dissipative hydrodynamic solutions is valid also in spheroidal and ellipsoidal, as well as rotating,  parametric solutions of the Navier-Stokes equations.  Thus we conjecture here, that the phenomena of approach to ``perfection" is a rather general phenomena. 
In the present manuscript, we have proven that the phenomena of approach to ``perfection", namely the asymtotic equality of a dissipative hydrodynamical solution with a perfect fluid solution exists and its domain of validity includes the non-relativistic, spherical fireballs that are governed by the Navier-Stokes equations. This domain of validity  of the asymptotically perfect fluidity of dissipative flows is to be studied and extended to other kind of fireball hydrodynamics in future, subsequent studies.

Thus we have shown that the approach to ``perfection" is not specific to relativistic kinematics~\cite{Csanad:2019lcl,Csorgo:2020iug}, but seems to be a more general phenomenon, valid for certain exact, 1+3 dimensional, spherically symmetric, parameteric  solutions of the Navier-Stokes equations. Further generalizations of these results are in progress, but will be discussed in separate manuscripts.

\section*{Acknowledgments}
It is our pleasure to thank M. Csanád, M. I. Nagy and W. A. Zajc for inspiring discussions during an early phase of this investigation.
We thank W. J. Metzger for a careful reading of our manuscript. Our research has been supported by the Hungarian Office for Research, Development and Innovation (NKFIH) under grants K133046, FK 123842 and FK 123959, and also via the project: Nanoplasmonic Laser
Inertial Fusion Research Laboratory (NKFIH-874-1/2020). Our research has also been partially supported by the Frankfurt Institute for Advanced
Studies, Germany, the E{\"o}tv{\"o}s Lor\'and Research Network (ELKH), Hungary, and by the Research Council of Norway, grant no. 255253.


\bibliography{NonRel-Navier-Stokes_spherical}

\end{document}